\documentclass[11pt,a4paper]{article}
\usepackage{jheppub}
\usepackage{color}
\usepackage[colorlinks=true,linkcolor=blue]{hyperref}
\usepackage[english]{babel}
\usepackage[utf8]{inputenc}
\usepackage[OT2,T1]{fontenc}
\usepackage{graphicx}
\usepackage{amsmath,amssymb,amsfonts,amsthm}
\usepackage{bm,bbm}
\usepackage{tikz,stackrel}
\usepackage{pstricks}
\usepackage{pifont} 
\usepackage{changepage}

\newcommand{\bc}{\begin{center}}
\newcommand{\ec}{\end{center}}
\newcommand{\be}{\begin{equation}}
\newcommand{\ee}{\end{equation}}
\newcommand{\ba}{\begin{eqnarray}}
\newcommand{\ea}{\end{eqnarray}}
\def\bs{\begin{subequations}}
\def\es{\end{subequations}}

\def\cA{\mathcal{A}}

\def\cV{\mathcal{V}}

\def\H{{\rm H}}

\def\B{\Box}
\newcommand{\Eq}[1]{(\ref{#1})}

\def\cob{\color{blue}}
\newcommand{\au}[2]{#1.~#2}
\newcommand{\book}[5]{\emph{#1}, #2, #3, #4 (#5)}
\newcommand{\books}[4]{\emph{#1}, #2, #3 (#4)}
\newcommand{\oarX}[1]{\href{http://arxiv.org/abs/#1}{{\ttfamily\cob arXiv:#1}}}
\newcommand{\arX}[1]{\href{http://arxiv.org/abs/#1}{{\ttfamily\cob arXiv:#1}}}
\newcommand{\doin}[6]{\href{http://dx.doi.org/#1}{{\cob {\it #2} {\bf #3 #4} (#6) #5}}}
\newcommand{\doinn}[5]{\href{http://dx.doi.org/#1}{{\cob {\it #2} {\bf #3} (#5) #4}}}
\newcommand{\doij}[5]{\href{http://dx.doi.org/#1}{{\cob {\it #2} {\bf #3} (#5) #4}}}

\newcommand{\ndoinn}[5]{\href{#1}{{\cob {\it #2} {\bf #3} (#5) #4}}}

\newcommand{\procm}[6]{in \emph{#1}, #2  eds., #3, #4, #5 (#6)}
\newcommand{\tia}[1]{\textit{#1},}
\def\rme{e}
\def\rmd{d}
\def\rmi{i}
\def\Re{{\rm Re}}

\newcounter{listcounter}

\newtheorem{thm}{Theorem}

\renewcommand{\qedsymbol}{$\blacksquare$}

\begin{document}

\title{Tree-level scattering amplitudes in nonlocal field theories}

\author[a]{Leonardo Modesto,}
\emailAdd{lmodesto@sustech.edu.cn}
\affiliation[a]{Department of Physics, Southern University of Science and Technology, Shenzhen 518055, China}

\author[b]{Gianluca Calcagni}
\emailAdd{g.calcagni@csic.es}
\affiliation[b]{Instituto de Estructura de la Materia, CSIC, Serrano 121, 28006 Madrid, Spain}

\abstract{We prove in two ways that, for a special class of nonlocal field theories consistent with linear and non-linear stability at the classical level, and with unitarity and super-renormalizability or finiteness at the quantum level, the $n$-point tree-level scattering amplitudes are the same as those of the underlying local theory. In particular, the $n$-point amplitudes of nonlocal gravity, with or without coupling to matter, are the same as for Einstein's general relativity.}

\preprint{\doij{10.1007/JHEP10(2021)169}{JHEP}{2110}{169}{2021} [\arX{2107.04558}]}

\maketitle



\section{Introduction}\label{Intro}

Recent years have witnessed the rise of a number of gravitational models beyond general relativity attempting to address the classical singularity problem and remove the perturbative quantum divergences of Einstein's theory \cite{Ori09,Fousp,Calcagni:2017sdq}. In particular, a general class of classical and quantum theories with nonlocal dynamics have attracted much attention and have become one of the most active fields of research in quantum gravity. Consider the action
\be\label{nlqg}
S=\frac{1}{2\kappa^2}\int\rmd^Dx\,\sqrt{|g|}\,\left[R+R\gamma_0(\B)\,R+R_{\mu\nu}\gamma_2(\B)\,R^{\mu\nu}+\cV(R,R_{\mu\nu})\right],
\ee
where $\kappa^2=8\pi G$, $G$ is Newton's constant, $g$ is the determinant of the metric (with signature $(-,+,\cdots,+)$), $R$ is the Ricci scalar, $R_{\mu\nu}$ is the Ricci tensor and $\gamma_{0,2}(\B)$ are analytic functions, called form factors, of the Laplace--Beltrami operator $\B=\nabla_\mu\nabla^\mu$. Nonlocal quantum gravity (NLQG) \cite{Kra87,Kuz89,Modesto:2011kw,BGKM,Modesto:2017sdr} is a set of theories which select specific form factors with exponential or asymptotically polynomial behaviour, which have especially benign properties that preserve unitarity and improve renormalizability. Renormalization can be further improved to finiteness for some choices of the potential-like local term $\cV$, depending on the topological dimension $D$ of spacetime \cite{MoRa1}. At the classical level, the theory has a well-defined problem of initial conditions \cite{CMN2,CMN3} and it admits stable Ricci-flat solutions, i.e., background metrics $g_{\mu\nu}$ such that $R_{\mu\nu}[g]=0$ \cite{Li:2015bqa,Calcagni:2017sov,Calcagni:2018pro,Briscese:2019rii}.\footnote{Other exact solutions can be found in \cite{Nascimento:2021bzb}.} Furthermore, theories with higher-order curvature terms usually carry at least one extra scalar degree of freedom dubbed scalaron or curvaton. However, this and other extra tensor modes do not propagate in NLQG on Minkowski spacetime \cite{CMN3}. Finally, some of the Ricci-flat solutions are singular (for instance, the Schwarzschild metric \cite{Calcagni:2017sov,Calcagni:2018pro}), which means that the classical nonlocal theory possesses, in principle, the classical black-hole singularities of general relativity. However, the finite version of the quantum theory is conformally invariant (because all beta functions vanish and so does the conformal anomaly) and conformal symmetry removes the singularities \cite{narlikar:1977nf,MoRa3}.

All these results are modified when optionally adding to (\ref{nlqg}) a $({\rm Riemann})^2$ term $R_{\mu\nu\rho\sigma}\gamma_4(\B)\,R^{\mu\nu\rho\sigma}$, which is not necessary for the renormalizability of the theory because $\gamma_4$ does not appear in the graviton propagator \cite{Kuz89,Modesto:2011kw,MoRa1}. In this case, Ricci-flat spacetimes are no longer exact solutions and black-hole singularities, if any, may be resolved without invoking conformal symmetry \cite{Frolov:1979tu,Frolov:1981mz,Tseytlin:1995uq,Frolov:2015bia,Frolov:2015bta,Frolov:2016pav,Frolov:2016xhq,Edholm:2016hbt,Cornell:2017irh,Buoninfante:2018xiw,Koshelev:2018hpt,Buoninfante:2018rlq,Buoninfante:2018stt,Giacchini:2018wlf,Burzilla:2020utr}.

A particular and somewhat surprising feature of NLQG is that the four-graviton scattering amplitude for $\gamma_4 =0$ is exactly the same \cite{Dona:2015tra} of Einstein's theory of gravity in vacuum \cite{DeWitt:1967uc,Berends:1974gk,Grisaru:1975bx}. Indeed, delicate cancellations among different channels take place when $\gamma_4(\B)=0$ and conspire together to obtain this result, valid also in local Stelle gravity \cite{Holdom:2021hlo}. An important implication is that 
no extra degrees of freedom appear at the quantum level \cite{Briscese:2018oyx}, consistently with the classical spectrum \cite{CMN3}.

These results are clear in the light of Anselmi's field redefinition \cite{Anselmi:2006yh}, which maps the nonlocal theory (with no $({\rm Riemann})^2$ term) to the Einstein--Hilbert action. Since the S-matrix is invariant under such field redefinitions with analytic nonlocal operators, one can conclude that all the tree-level amplitudes, including the 4-graviton amplitude found in \cite{Dona:2015tra}, are the same as in general relativity. 

However, one may wonder whether this conclusion holds also in the presence of matter. 
%
 This issue has been addressed recently in \cite{Modesto:2021ief}, where examples of nonlocal theories were given such that matter does not affect the stability properties of the local underlying local theory. 
%
More precisely, in that paper, a recipe was provided in order to construct nonlocal field theories starting from a local action and securing the following properties: linear and non-linear stability of all solutions of the local theory, super-renormalizability or finiteness at the quantum level, and perturbative unitarity at any loop order. 

We here will first recall the general action proposed in \cite{Modesto:2021ief} (section \ref{sec:gefre}), which has been already applied to the Higgs mechanism in nonlocal field theory \cite{Modesto:2021okr}, and afterwards we will prove in two different but equivalent ways that all the $n$-point tree-level scattering amplitudes of the nonlocal theory are identical to the amplitudes of the underlying local one. The first proof (section \ref{sec:redef}) is based on Anselmi's theorem \cite{Anselmi:2006yh} which was already implemented in nonlocal purely gravitational theories in \cite{Dona:2015tra}.\footnote{In the same paper \cite{Dona:2015tra}, it was shown that the scattering amplitudes in Stelle's theory and in Weyl conformal gravity are the same of the Einstein--Hilbert theory as long as we have only gravitons as external particles. When ghosts appear as external states, the amplitudes were computed in \cite{Johansson:2018ues}.} The second proof (section \ref{sec:pert}), which is a generalization of the results of \cite{Monteiro:2011pc} already applied to higher-derivative theories in \cite{Johansson:2018ues}, is based on the explicit construction of a unique correspondence between perturbative solutions and scattering amplitudes in the nonlocal theory.


\section{A general nonlocal framework with entire form factors}\label{sec:gefre}

Let us summarize the nonlocal theory proposed in \cite{Modesto:2021ief}. Here we use a compact DeWitt notation such that with the indices $i$, $j$ on fields we encode all Lorentz, group indices, and the spacetime dependence of the fields. Additionally, we assume that the field space is flat and we do not need to raise indices in sums there. The action in $D$ dimensions consists of the usual local Einstein's gravitational theory in the presence of matter extended with an operator quadratic in the Einstein's equations of motion (EoMs)\footnote{With an abuse of terminology, we will keep calling the objects $E_i$ EoMs.} $E_i=0$ between which it is inserted a nonlocal form factor $F_{ij}$ depending only on the Hessian operator $\Delta$ of the local theory. In a completely general notation, the action is
\ba
S(\Phi_i) &=& \int \rmd^D x \sqrt{|g|} \left[{\mathcal L}_{\rm loc} + E_i \, F_{ij}(\Delta) \, E_j + \cV(E)\right],\label{actiongen} \\
S_{{\rm loc}}  &=& \int \rmd^D x \sqrt{|g|} \, {\mathcal L}_{\rm loc} \, ,\qquad  \mathcal{L}_{\rm loc} = \frac{1}{2\kappa^2} R + \mathcal{L}_{\rm m}\,,\label{localT}\\
E_i(x) &:=& \frac{\delta S_{\rm loc}}{\delta \Phi_i(x)} \, , \\
\Delta_{ki} &:=& \frac{\delta E_i}{\delta \Phi_k} = \frac{\delta^2 S_{\rm loc}}{\delta \Phi_k \delta \Phi_i }\,,\label{EH}
\ea
where  $\Delta$ is the Hessian of the local theory, $\Phi_i=g_{\mu\nu},A_\mu,\phi,\dots$ is any field including the metric and matter fields and $F_{ij}(\Delta)(x,y)$ is a form factor related to an exponential entire function $\exp\H(\Delta)$ through the relation
\be
2 \Delta_{i k} F_{k j}(\Delta) \equiv \left[\rme^{\H(\Delta)}-1\right]_{ij}\,,\label{FF}
\ee
where repeated indices are summed over. Finally, the potential-like term $\cV(E)$ must be at least cubic in the local EoMs $E_i$, namely $\cV(E) = O(E^3)$. 

From the definition (\ref{FF}), the EoMs for the nonlocal theory turn out to be very simple:
\be
{\mathcal E}_k =  \left[\rme^{\H(\Delta)}\right]_{kj} \, E_j + O(E^2) = 0 \, .\label{LEOM}
\ee
Therefore, all the solutions of the local theory are solutions of the nonlocal theory, too, and, most importantly, we will show that the linear and non-linear perturbations of (\ref{LEOM}) satisfy the same EoMs as for the local theory. To be more explicit, given an exact background solution $E_k =0$ of the local theory, we can derive the EoMs for the perturbations expanding the fields, and then of the EoM, in a small dimensionless parameter $\epsilon$, i.e.,
\ba
\Phi_i &=& \sum_{n=0}^{\infty} \epsilon^n \Phi_i^{(n)}\,,\label{phin}\\
E_k(\Phi_i) &=&  \sum_{n=0}^{\infty} \epsilon^n E_k^{(n)}\,, \qquad \mathcal{{E}}_k(\Phi_i)=\sum_{n=0}^{\infty} \epsilon^n \mathcal{E}_k^{(n)}\,.  \label{ExpEpsilon}
\ea
Assuming that the fields $\Phi_i^{(0)}$ satisfy the local background EoMs
\be
E^{(0)}_k( \Phi_i^{(0)} ) =0 \,  ,\label{background}
\ee
it is extremely simple to prove a result which generalizes the theorems of \cite{BrMo,Briscese:2019rii}. In Appendix \ref{Proof}, we show that all perturbations (for gravity and matter) of the nonlocal theory satisfy the same EoMs of the perturbations in Einstein's gravity coupled to matter,
\be
{\mathcal{ {E}}^{(n)}_k}(\Phi_i^{(n)}) = 0 \quad  \Longrightarrow \quad  {E^{(n)}_k} (\Phi_i^{(n)}) = 0 \,\,\,\, \mbox{for} \,\,\,\, n>0 \,,\label{LeoTheorem}
\ee
where the label $n$ indicates the perturbative expansion of the tensors ${\mathcal{{E}}_k}$ and the EoM $E_k$ at the order $n$ in all the perturbations $\Phi_i^{(n)}$. The outcome (\ref{LeoTheorem}) is also an equivalent way to show tree-level unitarity, as we will see in section \ref{sec:pert}, while the Cutkosky rules \cite{Cutkosky:1960sp} and perturbative unitarity for nonlocal field theories have been considered in \cite{Briscese:2018oyx}. 

Now we are ready to prove the two theorems mentioned at the beginning of this introduction. 


\section{Field redefinition and scattering amplitudes}\label{sec:redef}

Since the action (\ref{actiongen}) consists of the local theory (\ref{localT}) augmented by nonlocal and analytic operators at least quadratic in the EoM $E_i$, the local and nonlocal theories are equivalent by means of an analytic field redefinition of the fields \cite{Anselmi:2006yh,Dona:2015tra,Koshelev:2017ebj,GiMo}. Let us review and adjust Anselmi's theorem \cite{Anselmi:2006yh,Dona:2015tra} to the theory (\ref{actiongen}). 

\begin{thm}
All the on-shell $n$-point tree-level amplitudes in the theory (\ref{actiongen}) are the same as the on-shell tree-level amplitudes of the theory (\ref{localT}).
\end{thm}
Let us assume to have two general weakly nonlocal or local action functionals $S(\Phi_i)$ and $S_{\rm loc}(\Phi_i)$ such that
%
%
\be
S(\Phi_k) = 
 S_{\rm loc}(\Phi_k) + E_i(\Phi_k) F_{i j}(\Phi_k)  E_j (\Phi_k) \, ,
 \label{AnselmiC}
\ee
where $F$ can contain derivative operators and ${E_i = \delta S_{{\rm loc}}/\delta \Phi_i}$ are the EoMs of the local theory. Hence, the statement of the theorem is that there exists a field redefinition 
\be
\Phi_i'  = \Phi_i + Q_{i j} E_j \, ,  \quad Q_{i j} = Q_{j i}, 
\label{FRQ}
\ee
such that, perturbatively in $F$, but to all orders in powers of $F$, we have the equivalence
\be
S(\Phi_k) = 
S_{\rm loc}(\Phi_k')  \,.
\label{FR}
\ee
Above, $Q_{ij}$ is a generic nonlocal operator acting linearly on $E_j$, 
and it is defined perturbatively in powers of the operator $F_{ij}$, namely, $Q_{ij}=F_{ij}(\Phi_k)+\ldots$.

{\color{black}{\bf Proof.}} 
 Let us consider the first order in the Taylor expansion for the functional $S_{\rm loc}(\Phi_k')$,
\be
S_{\rm loc}(\Phi_k') = S_{\rm loc}(\Phi_k + \delta \Phi_k) \simeq S_{\rm loc}(\Phi_k) + 
\frac{\delta S_{\rm loc}}{\delta \Phi_i} 
   \delta \Phi_i = 
 S_{\rm loc}(\Phi_k) + E_i   \, \delta \Phi_i \, .
\ee
If we can find a weakly nonlocal (analytic) expression for $\delta \Phi_i$ such that (note that the argument of the functionals $S$ and $S_{\rm loc}$ is now the same)
\be
S(\Phi_k) = 
S_{\rm loc}(\Phi_k) + E_i   \, \delta \Phi_i\,,
\ee
then there exists a field redefinition $\Phi_i \rightarrow \Phi_i'$ satisfying \eqref{FR}. Hence, the two actions $S(\Phi_k)$ and $S_{\rm loc}(\Phi_k')$ are tree-level equivalent. The general field redefinition (\ref{FRQ}) at any order in $F$ is given in \cite{Anselmi:2006yh} and the application of the theorem to (\ref{actiongen}) yields
\be
S_{{\rm loc}}(\Phi'_k) =  S_{\rm loc}(\Phi_k) + E_i(\Phi_k) F_{i j}(\Phi_k)  E_j (\Phi_k) = S(\Phi_k) 
\, .
\ee
Therefore, the nonlocal functional of the fields $\Phi_k$ coincides with the local functional of the fields $\Phi'_k$.

For the theory (\ref{actiongen}), the perturbative spectrum is the same of the local theory (\ref{localT}). Therefore, the Hilbert space of the asymptotic states of the nonlocal and local theories coincide. Therefore, by virtue of the field redefinition, the tree-level scattering processes are all and only the same. \qedsymbol

This result was shown in vacuum in \cite{Dona:2015tra} and we have generalized it here to a theory with matter. We will reach the same conclusion in section \ref{sec:pert} via a different route.


\section{Scattering amplitudes from the perturbative solutions of the EoM} \label{sec:pert}

Once established that all and only the perturbative solutions of the local theory (around a background that solves the EoMs of the local solution) are also solutions of the nonlocal theory (see section \ref{Intro}), we can focus on Minkowski spacetime and conclude that all the $n$-point scattering amplitudes of the local and nonlocal theories coincide.

A foundational result in local quantum field theory is that the solution of the classical equations of motion is the generating functional of tree graphs \cite{Nambu:1968rr,Boulware:1968zz}. This observations allows to find scattering amplitudes recursively \cite{Berends:1987me,Brown:1992ay} via a procedure which can be summarized the formula for the $n$-point scattering amplitude \cite{Monteiro:2011pc,Johansson:2018ues}
\be
\cA^{i_1 \dots i_n}_n(p_1, \dots, p_n) = \lim_{p_n^2 \rightarrow - m_n^2} \,
 \frac{1}{\rmi\tilde G(p_n)} \,
\frac{\delta^{n-1} \tilde{\Phi}^{(n-2)}_{i_n} (-p_n)}{\delta \tilde{\Phi}^{(0)}_{i_1}(p_1)  \cdots \delta \tilde{\Phi}^{(0)}_{i_{n-1}}(p_{n-1})},\label{AmplitudeG}
\ee
where $\tilde\Phi_i(p)$ is the Fourier transform of the generic field $\Phi_i(x)$, $\tilde G(p)$ is the tree-level Green's function, $\rmi\tilde G$ is the propagator, $\tilde{\Phi}^{(n-2)}_{i_n}$ is the solution at the $(n-2)$ perturbative order in a small-parameter expansion and $\tilde{\Phi}^{(0)}_{i}$ is the solution at zero order in the same expansion. For an ordinary quantum field theory with second-order kinetic operator, $\tilde G(p_n)=-1/p_n^2$ and one recovers the formula of \cite{Monteiro:2011pc}:
\be
\cA^{i_1 \dots i_n}_n(p_1, \dots, p_n) = \rmi\lim_{p_n^2 \rightarrow - m_n^2} \,
 p_n^2\,\frac{\delta^{n-1} \tilde{\Phi}^{(n-2)}_{i_n} (-p_n)}{\delta \tilde{\Phi}^{(0)}_{i_1}(p_1)  \cdots \delta \tilde{\Phi}^{(0)}_{i_{n-1}}(p_{n-1})}.\label{AmplitudeG0}
\ee
Notice that, for massless particles, the limit $p_n^2\to 0$ does not imply that $p_n^\mu\to 0$ for all $\mu=0,1,2,3$.


According to the proof in the Appendix \ref{Proof}, the perturbative solutions of (\ref{LEOM}) or (\ref{LEOM2}) are the same as for the local theory as long as we start from a background that solves the EoMs $E_i = 0$. Since Minkowski is one such background, we can apply directly \Eq{AmplitudeG} to the nonlocal theory (\ref{actiongen}) and get the $n$-point scattering amplitude, where $\tilde{\Phi}^{(n-2)}_{i_n}$ is the solution at the $(n-2)$-order in $\epsilon$ and $\tilde{\Phi}^{(0)}_{i}$ is the solution at zero order in the same expansion \Eq{ExpEpsilon}. Now, in nonlocal quantum gravity the form factor $\exp\H(\Delta)$ is such that, by construction, $\exp\H(-m^2)=1$ or at low momentum in the case of massless particles such as the graviton. Since the two-point 
 Green's function is always of the form $\tilde G(k)\simeq -\exp(-\H)/(k^2+m^2)$, no new poles are introduced in the spectrum of the theory and \Eq{AmplitudeG} reduces to \Eq{AmplitudeG0}. Therefore, we can conclude that 
\begin{thm}
All the $n$-point scattering amplitudes of the nonlocal theory (\ref{actiongen}) equal the amplitudes in the local theory (\ref{localT}). 
\end{thm}
For this result, it is crucial to have a benign type of nonlocality where the form factor is entire and $\exp\H\to 1$ on shell.

A final remark is in order. Let us focus on the weakly nonlocal term in the theory (\ref{actiongen}) with the asymptotically polynomial form factor typical of nonlocal quantum gravity \cite{Kuz89,Modesto:2011kw}. For the sake of simplicity, we only consider one field $\Phi$ and one cubic interaction. The leading contribution to the action in the ultraviolet regime reads
\be
 E \, F(\Box) \, E \sim  \Phi \Box^{\gamma} \Phi + \Phi^2 \Box^{\gamma} \Phi \, , \quad (\gamma > 4 \, , \,\, \gamma \in \mathbb{N})\, .
\ee
On the basis of the topological relation $N_I - N_V = L - 1$ ($N_V = N_I+1$ for $L=0$, $N_I$: number of internal lines, $N_V$: number of vertices, $L$: number of loops in a Feynman diagram), one might be induced to think that the tree-level amplitude diverges at most as
\be
\cA_n \sim \frac{p^{2 \gamma N_V}}{p^{2 \gamma N_I}} = \frac{p^{2 \gamma (N_I + 1)}}{p^{2 \gamma N_I}}= p^{2 \gamma} \, .
\ee
However, this naive counting fails for the special theories \Eq{actiongen} and also in higher-derivative theories where the external on-shell states that are supposed to interact are the same of the underling local two-derivative theories. On the other hand, more generally, in higher-derivative theories the spectrum of the theory usually contains other normal or ghost-like states and the amplitudes involving such asymptotic states will be usually non-zero.
%


\section{An explicit scalar-field example}\label{scafiex}

Let us consider now the example of a scalar field theory with a general potential $V(\phi)$. The action in $D=4$ dimensions reads
\ba
S(\Phi_i) &=& \int \rmd^4 x \sqrt{|g|} \left[ {\mathcal L}_{\rm loc} + E_\phi \, F(\Delta_\phi) 
\, 
E_\phi 
\right], \label{actionphi} \\
S_{{\rm loc}} &=& \int \rmd^4 x \sqrt{|g|} \, {\mathcal L}_{\rm loc} \, ,\qquad 
 \mathcal{L}_{\rm loc} = \frac{1}{2} \phi \Box \phi - V(\phi) \, , 
  \label{localTphi}\\
E_\phi &=& \frac{\delta S_{\rm loc} }{\delta \phi} = \Box \phi - V'(\phi) \,,\label{Ephi}
\\
\Delta_{\phi}&=&  \frac{ \delta E_\phi }{\delta \phi} 
= \frac{ \delta^2 S_{\rm loc} }{\delta \phi \, \delta \phi } = \Box  - V''(\phi)\,,\label{Deltaphi}\\
F(\Delta_\phi) &=& \frac{\rme^{\H(\Delta_\phi)}-1 }{2\Delta_\phi}\,.\label{Fphi}
\ea
The nonlocal EoM for the nonlocal scalar field theory is obtained taking the variation of the action (\ref{actionphi}) with respect to $\phi$, but the explicit variation is given in Appendix \ref{EoMphig}. The final result reads
\be\label{EoMphiAppe}
\rme^{\H(\Delta_\phi)} \, E_\phi + O(E_\phi^2)  = 0\,.
\ee
Therefore, according to the general proof in Appendix \ref{Proof}, the perturbative solutions of the nonlocal theory compatible with the background solutions of $E_\phi = 0$ are all and only the perturbative solutions of the local theory. Let us assume as usual that the exact background solution for $\phi$ is $\phi_{\rm B} = 0$, which trivially solves the EoM $E_\phi = 0$. This solution $\phi_{\rm B}$ would correspond to the Minkowski metric $\eta_{\mu\nu}$ in the case of gravity. 

To further simplify our example, we consider a scalar theory with only a cubic interaction,
\be
\mathcal{L}_{\rm loc} = \frac{1}{2} \phi \Box \phi - \frac{g}{3!} \phi^3 \, , \label{phi3}
\ee
whose local EoM reads
\be
\Box \phi - \frac{1}{2} g \phi^2 = 0 . \label{EoMphi3}
\ee
Introducing the Fourier transform for the field $\phi$ defined by:
\be
\phi(x) = \int \frac{\rmd^4 p}{(2 \pi)^4} \tilde{\phi}(p)\,\rme^{- \rmi p\cdot x} ,
\ee
the EoM turns into
\be
\int \frac{\rmd^4 k}{(2 \pi)^4} \, k^2 \tilde{\phi}(p)\, \rme^{- \rmi k\cdot x} + \frac{g}{2} \int \frac{\rmd^4 p_1}{(2 \pi)^4}
 \int \frac{\rmd^4 p_2}{(2 \pi)^4} \tilde{\phi}(p_1) \tilde{\phi}(p_2)\,  \rme^{- \rmi (p_1+p_2)\cdot x} = 0 \, .
\ee
Changing variable from $p_2$ to $k=p_1+p_2$, 
\ba
&& \int \frac{\rmd^4 k}{(2 \pi)^4} k^2 \tilde{\phi}(k)\, \rme^{- \rmi p\cdot x} + \frac{g}{2}
\int \frac{\rmd^4 k}{(2 \pi)^4} \int \frac{\rmd^4 p_1}{(2 \pi)^4} \tilde{\phi}(p_1) \tilde{\phi}(k - p_1)\,\rme^{-\rmi k\cdot x}= 0  \, ,  \nonumber \\ 
&& k^2 \tilde{\phi}(k) + \frac{g}{2} \int \frac{\rmd^4 p_1}{(2 \pi)^4} \tilde{\phi}(p_1) \tilde{\phi}(k - p_1) = 0 \, .
\label{Steps}
\ea
Finally, inserting the identity in terms of Dirac's delta in (\ref{Steps}), the equation of motion in momentum space reads
\be 
k^2 \tilde{\phi}(k) + \frac{g}{2} \int \frac{\rmd^4 p_1}{(2 \pi)^4} \int \frac{\rmd^4 p_2}{(2 \pi)^4}
\, \tilde{\phi}(p_1) \tilde{\phi}(p_2)  (2\pi)^4 \delta^4(k- p_1-p_2 ) = 0 \, . \label{EoMphi3k}
\ee
Now we make use of the perturbative solution of Eq.\ (\ref{EoMphi3k}) in momentum space:
\be
\tilde{\phi}(k) = \sum_{n = 0}^{+\infty} \tilde{\phi}^{(n)}(k) \, , \qquad \tilde{\phi}^{(n)}(k) \sim g^n \, .
\ee

As stated above, the full background solution is $\phi_{\rm B} = 0$. At zero order in the coupling $g$, namely $g^0$, the solution approximates to $\phi^{(0)}(x)$, which satisfies the plane wave equation with null momentum $k^2=k_\mu k^\mu=0$:
\be
\Box \phi^{(0)}(x) = 0 \quad \Longrightarrow \quad -k^2 \tilde{\phi}^{(0)}(k) = 0 
 \quad \Longrightarrow \quad  k^2 = 0\, .
\label{zero}
\ee
At order one in the coupling $g$, namely $g^1$, the solution of
\be
k^2 \tilde{\phi}^{(1)}(k) + \frac{g}{2} \int \frac{\rmd^4 p_1}{(2 \pi)^4} 
\int \frac{\rmd^4 p_2}{(2 \pi)^4}\,\tilde{\phi}^{(0)}(p_1) \tilde{\phi}^{(0)}(p_2) (2\pi)^4
 \delta^4(k- p_1-p_2 ) = 0 \,,\label{unoEqM}
\ee
is
\be
\tilde{\phi}^{(1)}(k) =  -\frac{g}{2} \int \frac{\rmd^4 p_1}{(2 \pi)^4} 
\int \frac{\rmd^4 p_2}{(2 \pi)^4}\,\frac{1}{k^2} \, \tilde{\phi}^{(0)}(p_1) \tilde{\phi}^{(0)}(p_2) 
(2\pi)^4 \delta^4(k- p_1-p_2 )\,.\label{uno}
\ee
At second order in $g$, namely $g^2$,
\ba
\tilde{\phi}^{(2)}(k) \!&=&\! -\frac{g}{2} \int \frac{\rmd^4 p_1}{(2 \pi)^4} 
\int \frac{\rmd^4 p_2}{(2 \pi)^4}\,\frac{1}{k^2}\left[
\tilde{\phi}^{(0)}(p_1) \tilde{\phi}^{(1)}(p_2)  
+ \tilde{\phi}^{(1)}(p_1) \tilde{\phi}^{(0)}(p_2)  \right](2 \pi)^4 \delta^4(k- p_1-p_2 ) \nonumber \\
&=& -g \int \frac{\rmd^4 p_1}{(2 \pi)^4}\int \frac{\rmd^4 p_2}{(2 \pi)^4}\,\frac{1}{k^2} \, \tilde{\phi}^{(0)}(p_1) \, \tilde{\phi}^{(1)}(p_2)\,(2 \pi)^4\delta^4(k- p_1-p_2 )\,,\label{due}
\ea
which is symmetric exchanging $p_1$ and $p_2$. 

Now we replace the first-order solution \Eq{uno} into \Eq{due}:
\ba
\tilde{\phi}^{(2)}(k) & = & -g \int  \frac{\rmd^4 p_1}{(2 \pi)^4} \int \rmd^4 p_2\,\frac{1}{k^2}\nonumber\\
&&\qquad\times
\left[-\tilde{\phi}^{(0)}(p_1) 
\frac{g}{2} \int \frac{d^4 k_1}{(2 \pi)^4} 
\int \rmd^4 k_2\,\frac{1}{p_2^2} \, 
\tilde{\phi}^{(0)}(k_1) \tilde{\phi}^{(0)}(k_2)  \delta^4(p_2- k_1-k_2 )
\right]\nonumber\\
&& \qquad\times\delta^4(k- p_1-p_2) \nonumber \\
& = & 
\frac{g^2}{2} \! \int  \! \frac{\rmd^4 p_1}{(2 \pi)^4} \, \frac{1}{k^2}\nonumber\\
&&\qquad\times \left[\tilde{\phi}^{(0)}(p_1) 
 \int \frac{\rmd^4 k_1}{(2 \pi)^4} 
\int \rmd^4 k_2\, \frac{1}{(k_1 + k_2)^2} \, 
\tilde{\phi}^{(0)}(k_1) \tilde{\phi}^{(0)}(k_2) \right]\nonumber\\
&&\qquad\times\delta^4(k- p_1- k_1 - k_2 )\,,\label{dueB}
\ea
where we integrated in $p_2$. Making the replacements $k_1 \rightarrow p_2$ and $k_2 \rightarrow p_3$,
\ba
\tilde{\phi}^{(2)}(k) &=&
\frac{g^2}{2} \! \int  \! \frac{\rmd^4 p_1}{(2 \pi)^4} 
 \int \frac{\rmd^4 p_2}{(2 \pi)^4} 
\int \frac{\rmd^4 p_3}{(2 \pi)^4} 
\frac{1}{k^2} 
\frac{1}{(p_2 + p_3)^2} \,\tilde{\phi}^{(0)}(p_1)\, \tilde{\phi}^{(0)}(p_2) \,\tilde{\phi}^{(0)}(p_3)\nonumber\\
&&\qquad\qquad\qquad\times(2 \pi)^4 \delta^4(k- p_1- p_2 - p_3 ) \,.\label{dueC}
\ea

We can now compute three-point and four-point scattering amplitudes. For that purpose, we recall that the functional variation in momentum space is decorated with a $(2\pi)^4$ from the Fourier transform:
\be
\frac{\delta\tilde\phi(p)}{\delta\tilde\phi(p')}=(2\pi)^4\delta^4(p-p')\,.
\ee 
Let us start with the {\em 3-point scattering amplitude}. Replacing the solution at the first order in $g$ in formula \Eq{AmplitudeG} for $n=3$ we get:
\ba
\cA_3 (p_1, p_2, p_3) & = & \rmi\lim_{p_3^2 \rightarrow 0} \,\rme^{\H(-p_3^2)} p_3^2 \,\frac{\delta^{2} \tilde{\phi}^{(1)}( - p_3)}{\delta \tilde{\phi}^{(0)}(p_1)  \,  \tilde{\phi}^{(0)}(p_{2}) }\nonumber \\
& = & \rmi\lim_{p_3^2 \rightarrow 0} \, p_3^2 \, \frac{\delta^{2} \left[-\frac{g}{2} \int \frac{\rmd^4 k_1}{(2 \pi)^4} \int \rmd^4 k_2\, 
\frac{1}{(-p_3)^2} \, \tilde{\phi}^{(0)}(k_1) \tilde{\phi}^{(0)}(k_2)  \delta^4(k_1+k_2+p_3)\right]}{\delta \tilde{\phi}^{(0)}(p_1)  \,  \tilde{\phi}^{(0)}(p_{2}) } \nonumber \\
& = & -\rmi \lim_{p_3^2 \rightarrow 0} \, \frac{g}{2} (2\pi)^4\int\rmd^4 k_1\int \rmd^4 k_2\,\delta^4(k_1+k_2+p_3)\nonumber\\
&&\qquad\qquad\times \left[ \delta^4(p_1-k_1) \, \delta^4(p_2-k_2) + \delta^4(p_1-k_2) \, \delta^4(p_2-k_1) \right]  \nonumber \\
& = & -\rmi\,g(2\pi)^4\, \delta^4(p_1+p_2+p_3) \,,\label{Amplitude3}
\ea
which coincides with the well-known result in local quantum field theory.

The computation of the {\em 4-point scattering amplitude} is slightly more tedious and involved. From formula \Eq{AmplitudeG} for $n=4$ the expression to evaluate is
\be  
\cA_4 (p_1, \dots, p_4) = \rmi  \lim_{p_4^2 \rightarrow 0} \,\rme^{\H(-p_4^2)} p_4^2 \,\frac{\delta^{3} \tilde{\phi}^{(2)} ( - p_4 )}{\delta \tilde{\phi}^{(0)}(p_1) \, \delta \tilde{\phi}^{(0)}(p_2)  \,   \delta \tilde{\phi}^{(0)}(p_{3}) } .\label{Amplitude4}
\ee
The solution at second order in $g$ and evaluated in $k = - p_4$ reads
\ba
\tilde{\phi}^{(2)}(- p_4) &=&\frac{g^2}{2} \! \int  \! \frac{\rmd^4 k_1}{(2 \pi)^4} \int \frac{\rmd^4 k_2}{(2 \pi)^4} 
\int \rmd^4 k_3 \frac{1}{p_4^2} \frac{1}{(k_2 + k_3)^2} \, \tilde{\phi}^{(0)}(k_1)\, \tilde{\phi}^{(0)}(k_2) \,\tilde{\phi}^{(0)}(k_3)\nonumber\\
&&\qquad\qquad\times \delta^4(k_1+k_2 + k_3 +p_4)\,,\label{dueD}
\ea
In order to get the final $4-$points amplitude we have to evaluate the functional derivative in (\ref{Amplitude4}), which yields
\ba
\hspace{-1cm}&&\frac{1}{(2\pi)^{12}}\frac{\delta^{3} [\tilde{\phi}^{(0)}(k_1)\, \tilde{\phi}^{(0)}(k_2) \,\tilde{\phi}^{(0)}(k_3)]}{\delta \tilde{\phi}^{(0)}(p_1) \, \delta \tilde{\phi}^{(0)}(p_2)  \,   \delta \tilde{\phi}^{(0)}(p_{3}) }\nonumber\\  
\hspace{-1cm}&&\qquad\qquad= \delta(p_1-k_1) \, \delta(p_2-k_2) \, \delta(p_3-k_3)+\delta(p_1-k_2) \, \delta(p_2-k_3) \, \delta(p_3-k_1)\nonumber \\
\hspace{-1cm}&&\qquad\qquad\quad+\delta(p_1-k_3) \, \delta(p_2-k_1) \, \delta(p_3-k_2)+\delta(p_1-k_3) \, \delta(p_2-k_2)\,\delta(p_3-k_1)\nonumber \\
\hspace{-1cm}&&\qquad\qquad\quad+\delta(p_1-k_2) \, \delta(p_2-k_3) \, \delta(p_3-k_1)+\delta(p_1-k_3) \, \delta(p_2-k_1) \, \delta(p_3-k_2)\, .\label{3var}
\ea
Replacing (\ref{dueD}) into (\ref{Amplitude4}), using (\ref{3var}) and integrating in $k_1,k_2,k_3$, the amplitude finally reads
\be
\cA_4 (p_1, \dots, p_4) = \rmi\,g^2\left[ \frac{1}{(p_1+p_2)^2} + \frac{1}{(p_2+p_3)^2} + \frac{1}{(p_1+p_3)^2}\right] (2\pi)^4\delta^4(p_1+p_2+p_3+p_4)\,,
\ee
again a well-known result in local quantum field theory \cite{Sre07}. 


\section{Conclusions}

We have proved in two different ways that all the tree-level scattering amplitudes of the special class of theories defined in \Eq{actiongen} are the same of the underlying local theory \Eq{localT}. The first proof was based on a general field redefinition theorem that apply to a more general class of theories (section \ref{sec:redef}), while the second proof was constructive and based on a general algorithm that provides a correspondence between perturbative solutions of the classical equations of motion and tree-level scattering amplitudes (section \ref{sec:pert}). As a particular example, we applied the theorem to a scalar field theory with cubic interaction. 
These results equally apply to the NLQG \Eq{actiongen} because of the EoM \Eq{EoMphiAppe} whose perturbative solutions are the same of the Einstein's theory around any background spacetime that solves the Einstein's local EoM.


The class of models discussed here does not encompass all possible nonlocal ghost-free theories of gravitation, since it leaves out the cases with non-entire form factors, such as theories with a minimal length \cite{Padmanabhan:1996ap}, with a minimal proper time in Schwinger parametrization \cite{Abel:2019ufz,Abel:2019zou} or with fractional operators \cite{mf2}. On the other hand, the present work has been motivated by the need to clarify the topic of scattering amplitudes in nonlocal gravity \cite{Kra87,Kuz89,Modesto:2011kw,Modesto:2017sdr,MoRa1,MoRa2,Modesto:2015foa} but it actually has a very general applicability and validity. In particular, the same analysis applies also to local higher-derivative theories and it is consistent with what we already know. In the case of Lee--Wick theories, such as the quantum gravity proposal of \cite{Modesto:2015ozb,Modesto:2016ofr}, complex conjugate ghosts appear, but they can be consistently removed from the asymptotic spectrum \cite{Modesto:2015ozb} implementing the Anselmi-Piva prescription (and not the Feynman prescription) \cite{Anselmi:2017yux,Anselmi:2017lia,Ans18}. 
Indeed, the latter prescription guarantees that, in the loop amplitudes, ghosts are not restored consistently with the fact that they cannot go on-shell. The same prescription works also for more general higher-derivative theories \cite{Asorey:1996hz}, which, therefore, turns out to be unitary too. 

As a future refinement of the present results, we mention the possibility to find another proof via the world-line approach of \cite{Abel:2019ufz}. There, the propagator of a given scalar theory was expressed in the Schwinger representation where the integration over the proper-time parameter $s$ was performed with a non-trivial measure for $s$ and a minimal value $s_*$. If the propagator of the corresponding local scalar theory is obtained in the limit $s_*\to 0$, then, the reasoning goes, the tree-level scattering amplitudes of the nonlocal theory are the same as in the local theory, since the amplitudes of asymptotic in-out states should not be affected by the short-scale behaviour. Similar considerations could apply also to nonlocal quantum gravity. The graviton propagator has the form $\sim \exp[-\H(-k^2)]/k^2$, where $\H$ is an entire function. This admits a Schwinger-like representation of the form
\be
\frac{\rme^{-s_*\H}}{k^2}=\frac{\H}{k^2}\int_{s_*}^{+\infty}\rmd s\,\rme^{-s\H}\,,
\ee
valid for $\Re\,\H>0$. When $s_*=1$, one obtains the nonlocal propagator of the theory. When $s_*=0$, indeed one recovers the local propagator, for any $\H$. Although this representation does not recast all the derivative (momentum) dependence in integral form, the overall factor $\H/k^2$ does not have any pole because $\H$ has a trivial kernel, i.e., $\H(0)=0$ in this class of theories. Therefore, it admits a series expansion $\H(-k^2)=\sum_{l=1}^{\infty}c_l (k^2)^l$ with no zero mode and, consequently, $\H/k^2=c_1+c_2k^2+\dots$. Thus, the limit
\be
\lim_{k^2\to 0}\frac{\H(-k^2)}{k^2}=c_1
\ee
is finite. The details of this reasoning should be filled up rigorously but we do not expect major complications. However, we note that, in general, it is not sufficient to prove that the propagator does not display any other pole besides the usual ones because scattering amplitudes also involve the vertices and these may carry singularities such as single poles, double poles or branch cuts. 


\section*{Acknowledgments}

G.C.\ is supported by the I+D grants FIS2017-86497-C2-2-P and PID2020-118159GB-C41 of the Spanish Ministry of Science and Innovation.



\appendix


\section{Linear and nonlinear stability in the nonlocal field theory (\ref{actiongen})}\label{Proof}

In order to prove that the linear and non-linear stability of the nonlocal theory (\ref{actiongen}) are the same of the local theory, we have to expand perturbatively in $\epsilon$ (see Eq.\ (\ref{ExpEpsilon})) the EoM (\ref{LEOM}), which we write in vectorial notation:
\be
{\mathbf{\mathcal{E}}} =\mathbf{\rme^{\H(\Delta)} \, E} + O(\mathbf{E^2}) = 0\,.\label{LEOM2}
\ee
Since we want to study the stability of exact solutions of the Einstein's theory coupled to matter, we assume to expand around a metric consistent with  (\ref{background}), ${\bf E^{(0)}} = 0$.

Hence, at zero order in $\epsilon$, we have
\be 
\mathbf{\rme^{\H^{(0)}(\Delta)} \, E^{(0)}} + O(\mathbf{E^{(0) 2}}) = 0 \,, \label{LEOMzero}
\ee
which is satisfied because, by hypothesis, the background satisfies ${\bf E^{(0)}} = 0$. 

At the first order $\epsilon^1$, we get
\ba
&&\mathbf{\rme^{\H^{(1)}(\Delta)} \, E^{(0)}} + \mathbf{\rme^{\H^{(0)}(\Delta) } \, E^{(1)}} 
+ O(\mathbf{E^{(0)} E^{(1)}}) = 0 \nonumber\\
&& \qquad \Longrightarrow \quad  {\bf E^{(1)}} = 0 
\quad 
\Longrightarrow \quad \Phi_i^{(1)} = \Phi_i^{(1)} \! \left(\Phi_j^{(0)} \right)\,,\label{LEOMuno}
\ea
where we used the solution ${\bf E^{(0)}} = 0$ at the previous perturbative order. 

At the second order $\epsilon^2$, we get
\ba
&& \mathbf{\rme^{\H^{(2)}(\Delta) } \, E^{(0)}} + \mathbf{\rme^{\H^{(1)}(\Delta)} \, E^{(1)}} + \mathbf{\rme^{\H^{(0)}(\Delta) } \, E^{(2)}} + O(\mathbf{E^{(1)} E^{(1)}}) + O(\mathbf{E^{(2)} E^{(0)}}) = 0 \nonumber\\
&&\qquad \Longrightarrow \quad  {\bf E^{(2)}} = 0  \nonumber\\
&&\qquad \Longrightarrow \quad \Phi_i^{(2)} = \Phi_i^{(2)} \! \left(  \Phi_j^{(0)} ,  \Phi_k^{(1)} \right)= \Phi_i^{(2)} \! \left(  \Phi_j^{(0)} ,   \Phi_k^{(1)} \! \left(  \Phi_l^{(0)} \right) \right), \label{LEOMdue}
\ea
where we used ${\bf E^{(0)}} = 0$ and ${\bf E^{(1)}} = 0$. 

Finally, by induction, at the order $\epsilon^n$ we obtain
\ba
&& \mathbf{\rme^{\H^{(n)}(\Delta) } \, E^{(0)}}
+ \mathbf{\rme^{\H^{(n-1)}(\Delta) } \, E^{(1)}} 
+ \mathbf{\rme^{\H^{(n-2)}(\Delta) } \, E^{(2)}} + \dots 
+ \mathbf{\rme^{\H^{(0)}(\Delta) } \, E^{(n)}} + \nonumber \\ 
&& + O(\mathbf{E^{(n)} E^{(0)}}) 
+ O(\mathbf{E^{(n-1)} E^{(1)}}) + \dots
+ O(\mathbf{E^{(1)} E^{(n-1)}}) 
+ O(\mathbf{E^{(0)} E^{(n)}}) 
= 0 \nonumber\\
&&\qquad \Longrightarrow \quad  {\bf E^{(n)}} = 0 \nonumber\\
&&\qquad \Longrightarrow \quad \Phi_i^{(n)} 
= \Phi_{i_n}^{(n)} \! \left(  \Phi_{i_1}^{(0)} ,  
\dots , \Phi_{i_{n-1}}^{(n-1)} \right)
= \Phi_{i_n}^{(n)} \! \left(  \Phi_{i_1}^{(0)} ,  
\dots , \Phi_{i_{n-1}}^{(n-1)} \left(    \Phi_{i_{n-2}}^{(n-2)}, \dots , \Phi_{i_1}^{(0)}     \right)\right),
\nonumber\\\label{LEOMn}
\ea
where we used ${\bf E^{(0)}} = 0$, ${\bf E^{(1)}} = 0$, \dots, ${\bf E^{(n-1)}} = 0$. 

Therefore,
\be
\mathbf{\mathcal{E}^{(n)}} = 0 \quad \Longrightarrow \quad \mathbf{{E}^{(n)}} = 0 \, .
\ee
Since this condition is not only sufficient but also necessary (whenever $E=0$, also $\mathcal{E}=0$), we conclude that
\be
\mathbf{\mathcal{E}^{(n)}} = 0 \quad \Longleftrightarrow \quad \mathbf{{E}^{(n)}} = 0 \, .
\ee

It deserves to be notice that the exponential form factor always appear at the zero order in the $\epsilon-$expansion, namely ${\bf \exp H^{(0)}}$, at any step of the proof, since higher-order terms ${\bf \exp H^{(1)}}$, ${\bf \exp H^{(2)}}$, \dots, are always multiplied by a quantity which vanishes on the background.


\section{Equations of motion for a nonlocal scalar field theory} \label{EoMphig}

The nonlocal EoM for the nonlocal scalar field theory (\ref{actionphi}) is obtained taking the variation of the action:
\ba
\delta S & =  & \int \rmd^4 x \left[ \delta \phi \, E_\phi  + \left(\frac{\delta E_\phi}{\delta \phi}   \delta \phi \right)\, F(\Delta_\phi) \, E_\phi 
+ E_\phi \, F(\Delta_\phi) \,  \left(\frac{\delta E_\phi}{\delta \phi}   \delta \phi \right) + O(E_\phi^2) \right] 
\nonumber \\
& = & 
 \int \rmd^4 x \left\{  \delta \phi \, E_\phi  + \left[(\Box  - V'') \delta\phi \right] F(\Delta_\phi) \, E_\phi
+ E_\phi \, F(\Delta_\phi)    \left[(\Box -V'')\delta\phi \right] + O(E_\phi^2) \right\}.\nonumber\\\label{vari}
\ea
Integrating by parts and using the definitions (\ref{Deltaphi}) and (\ref{Fphi}), the variation (\ref{vari}) turns into
\ba
\delta S &  = & 
 \int \rmd^4 x \,  \delta \phi \left[  E_\phi  + 2 \left[(\Box  - V'')\right] F(\Delta_\phi) \, E_\phi
 + O(E_\phi^2) \right] \nonumber \\
 & = &
 \int \rmd^4 x \, \delta \phi \left[E_\phi  + 2 \Delta_\phi F(\Delta_\phi) \, E_\phi + O(E_\phi^2)   \right]  \nonumber \\
 & = &
 \int \rmd^4 x \, \delta \phi \left[  E_\phi  + 2 \Delta_\phi \frac{\rme^{\H(\Delta_\phi)} - 1 }{2 \Delta_\phi} \, E_\phi
 + O(E_\phi^2)   \right] \nonumber \\
 &=& \int \rmd^4 x \,  \delta \phi \left[\rme^{\H(\Delta_\phi)} \, E_\phi + O(E_\phi^2)   \right] ,\label{vari2}
\ea
implying that the EoM is (\ref{EoMphiAppe}).


\end{document}